\documentstyle[12pt]{article}
\newfont{\feff}{cmti10}
\begin{document}

\title{Scaling Exponents of 
Strong Turbulence in the Eddy Viscosity Approximation}

\author{ Victor Yakhot\\
Program in Applied and Computational Mathematics\\
Princeton University}

\maketitle

\begin{abstract}

A dynamic equation  for  velocity structure functions $S_{n}(r)=<(u(x)-u(x'))^{n}>$
in strong turbulence is derived in the one-loop (eddy viscosity) approximation. 
This homogemeous differential equation yields scaling exponents $\xi_{n}$ in 
the relations $S_{n}(r)\propto r^{\xi_{n}}$ which are in a very good 
agreement with experimental data.

\end{abstract}

\noindent 
Kolmogorov's (1941) relation  for the third-order structure function which is in 
the limit of zero viscosity $\nu_{0}\rightarrow 0$, 

$$S_{3}=<(u(2)-u(1))^{3}>=-\frac{4}{5}{\cal E}r\eqno(1)$$

\noindent where ${\cal E}=\nu\overline{(\partial_{i} u_{j})^{2}}$ is the mean dissipation rate of 
turbulent kinetic energy, $u(i)\equiv u({\bf x_{i}})$ is the value of the $x$-component of the velocity 
field at
the point ${\bf x}$, and $r={\bf n\cdot(x_{2}-x_{1})}$  is the value of the displacement along
the $x$-axis (${\bf n}$ is the unit vector in the $x$-direction) [1].  Applying 
dimensional 
considerations to this dynamic relation, Kolmogorov also made the  prediction:

$$S_{2}=<(u(2)-u(1))^{2}>\propto {\cal E}^{\frac{2}{3}}r^{\frac{2}{3}}\eqno(2)$$

\noindent leading to the celebrated Kolmogorov energy spectrum:

$$E(k)=C_{K}{\cal E}^{\frac{2}{3}}k^{-\xi}\eqno(3)$$

\noindent with $\xi=5/3$. Kolmogorov went even further by generalizing (2) to 
the structure function $S_{n}(r)$ 
of  arbitrary order $n$:

$$S_{n}(r)=<(u(2)-u(1))^{n}>=A_{n}({\cal E}r)^{\frac{n}{3}}\eqno(4)$$

Experimental investigations have supported the relation (2) with good accuracy, 
giving for the exponent of the second-order structure functions $\xi_{2}\approx 
1.66-1.75$.  At the same time substantial deviations of $\xi_{n}$ from the K41 values
$\xi_{n}=n/3$ from  $n>3$ have also been observed.  It is interesting that no data on the 
scaling exponents of the structure functions $S_{n}$ with $n<1$ have yet been reported.

Ever since Kolmogorov's work,  dimensional considerations were  the most 
dominant  way of evaluating the scaling exponents. This is why until  recently 
no progress in understanding the intermittency of strong turbulence was  achieved.
In 1994 R. H. Kraichnan,  considering the problem of a passive scalar advection 
in a rapidly-
changing- in- time random velocity field, realized that  anomalies,  coming from 
the dissipation terms in the equation of motion,  result in  a  homogeneous differential
equation for $S_{n}(r)$,  leading to an  algebraic form of the scalar structure functions 
$S_{n}(r)\propto r^{\xi_{n}}$ with  
scaling exponents  $\xi_{n}$ determined
 by the numerical values of the coefficients
in the equation [2].  This is why the scaling exponents  $\xi_{n}$ cannot be obtained on 
the basis  of dimensional considerations. Later,    groundbreaking works by
Gawedzkii and Kupiainen [3 ], Chertkov  et. al.  [4 ] proved that in the vicinity of the
gaussian limits  the scaling of  scalar structure functions is determined by the zero
modes of the homogeneous differential equations for $S_{n}(r)$,  which for 
this problem can be written explicitly [5  ], [ 6 ]. The results of Refs. [ 3]- [4 ]
have been confirmed in ref. [7 ] using a different approach. At about the same
time Shraiman and Siggia [8 ],  considering a model of a passive scalar 
based on  the concept of eddy diffusivity, showed that zero modes play the most 
important part in determination of the scaling exponents of  structure functions.

Since Richardson's  (1926) work,  effective transport coefficients 
(eddy viscosity and eddy diffusivity), though not
rigorously justified,  have been used for the quantitative description of 
transport phenomena in strongly  turbulent engineering flows. It is safe to state that 
by now
these concepts have evolved into an engineering tool widely used  for design purposes 
throughout mechanical engineering.  The idea behind the method is simple [9], [10].
Consider the Navier-Stokes equations for an incompressible fluid:

$$u_{it} +{\bf u\cdot \nabla}u_{i}
=S_{i} -\nabla_{i}p +\nu_{0}\nabla^{2}u_{i}\eqno( 5)$$

$${\bf \nabla\cdot u}=0$$

\noindent where ${\bf S(x,t)}$ (${\bf \nabla \cdot S}=0$) is a large-scale source 
function.   Using the incompressibility condition the pressure $p$ can be expressed 
in terms of the velocity field.  It is useful  to write  
the equation for the Fourier-transform ${\bf u(k,\omega)}$:

$$-i\omega u_{l}({\bf k},\omega)+
\frac{i}{2}P_{lmn}({\bf k})\int d{\bf q}d\Omega u_{m}({\bf q},\omega)u_{n}({\bf k-q}, \omega-\Omega)
=S_{l}({\bf k},\omega)-
\nu_{0} k^{2}u_{l}({\bf k},\omega)$$															

\noindent where the projection operator $P_{lmn}({\bf k})$ is:

$$P_{lmn}({\bf k})=k_{m}P_{ln}({\bf k})+k_{n}P_{lm}({\bf k})$$

\noindent and  $P_{ij}=\delta_{ij}-\frac{k_{i}k_{j}}{k^{2}}$.
In the eddy viscosity approximation the effects of the small scales ($q>>k$)
on the modes ${\bf u(k)}$ is represented in terms of the eddy viscosity [10 ] 
(we neglect the small eddy noise contribution):

$$-i\omega u_{l}({\bf k},\omega)+
\frac{i}{2}P_{lmn}({\bf k})\int d{\bf q}d\Omega u_{m}({\bf q},\omega)u_{n}({\bf k-q}, \omega-\Omega)\approx
 S_{l}-\Gamma  k^{2+a}u_{l}({\bf k},\omega)-
\nu_{0} k^{2}u_{l}({\bf k},\omega)\eqno(6)$$

\noindent where 

$$a=-\frac{2+\xi_{2}}{2}\eqno(7)$$

The integration  is carried out over the interval: $0<q<k$; ~$-\infty<\omega<\infty$.
This means that this equation describes  the velocity field averaged over small-scale  
fluctuations with $q>k$.  
This result 
must be 
considered as a general model for the  velocity field $u_{l}({\bf k},\omega)$  where
the effects of the small- scale fluctuations  are accounted  for by  the
eddy viscosity with $\nu(k)\propto \Gamma k^{a}$.  
In principle, this equation must be solved subject to 
initial and boundary
conditions for each mode $u({\bf k},t)$.  A similar model has recently been used 
for evaluation of the anomalous scaling in 
the problem of a passive scalar advected by a random velocity field by 
Shraiman and Siggia [8 ].  The role of the large scale motions in the dynamics of 
the small-scale 
velocity fluctuations, described by the non-linear contribution to (6),  
is two-fold.
First,  the large-scale structures kinematically transfer ( sweep ) the small-scale 
fluctuations. This process does not lead to the energy redistribution. Secondly,
the large- scale motions serve as an energy source for the small-scale dynamics  
which can be accounted for by the random forcing function ${\bf f}$ which is 
a complex functional of the velocity field.  Eventually,  we will be 
 interested in the equation 
of motion for the structure functions of velocity differences $\Delta u=u(2)-u(1)$, 
for which  sweeping  is not too important. 
This allows us to write (6)
as  a simple Langevin-like equation for ${\bf u(k)}$: 

$$-i\omega u_{l}({\bf k},\omega)\approx
 S_{l}+f_{l}-\Gamma  k^{2+a}u_{l}({\bf k},\omega)-
\nu_{0} k^{2}u_{l}({\bf k},\omega)\eqno(8)$$

In this approximation the equation for the structure function $S_{n}(r)$ can be 
written
readily in the three-dimensional case:

$$\frac{1}{r^{2}}\frac{\partial}{\partial r} r^{2-a}
\frac{\partial S_{n}}{\partial r}
=nD\eqno(9 )$$

\noindent where 

$$D=<(\Delta S+\Delta f+ \nu_{0}(\nabla^{2}_{2}u(2)-\nabla^{2}_{1}u(1)))(\Delta u)^{n-1}>$$

\noindent We are interested in the behaviour of the structure functions for small values $r/L<<1$.
Since the source $S$ is assumed to act at  large scales only we neglect 
it in what follows.   
As in the problem of a passive  scalar,   the derivation of the  $D$-term is a 
difficult task.  Here the problem is even harder  since we do not know much
about the effective energy source ${\bf f}$. It is clear, however, that the 
eddy viscosity approximation can be accurate only for $S_{n}(r)$ with the
 relatively small $n>0$.  Thus, the values of $S_{n}$ are dominated by the part of 
the probability density $P(\Delta u, r)$ where $\Delta u \approx (\Delta u)_{rms}$.  We 
assume:

$$D=f(r)S_{n}(r)\eqno(10)$$

\noindent with $f(r)$ independent on $n$. A  more detailed argument leading 
to (10) will be presented below where it will be shown that (10) is consistent 
with the eddy viscosity approximation. The expression (10) is not dissimilar 
to the anzatz introduced by Kraichnan in his  theory of a passive scalar [2]. The 
function $f(r)$ is fixed by the relation (1).  Introducing the new variable

$$r=\frac{4}{5}{\cal E}|x_{1}-x_{2}|\eqno(11)$$

\noindent so that 

$$S_{3}(r)=r\eqno(12)$$

\noindent we have 

$$\frac{1}{r^{2}}\frac{\partial}{\partial r} r^{2-a}
\frac{\partial S_{n}}{\partial r}
=\frac{n}{3}(2-a)r^{-2-a}S_{n}\eqno(13)$$

\noindent One can see that the relation (12) is satisfied by this equation. 
The corresponding equation for the probability density $P(\Delta u,r)\equiv P(U,r)$ is:

$$\frac{1}{r^{2}}\frac{\partial}{\partial r} r^{2-a}
\frac{\partial P(U,r)}{\partial r}
=-\frac{(2-a)}{3}
r^{-2-a}\frac{\partial }{\partial U}U~P(U,r)$$

The equation for the exponents $\xi_{n}$ following from (13) is:

$$\xi_{n}(1+\frac{2+\xi_{2}}{2}+\xi_{n})-\frac{n}{3}(2+\frac{2+\xi_{2}}{2})=0\eqno(14)$$. 

Thus, we see that all exponents $\xi_{n}$ are determined in terms of $\xi_{2}$.
Substituting $n=2$ into (14) we derive $\xi_{2}=0.7257$. Using this value we have:

$$\xi_{n}\approx -1.1815~+~(1.3958+1.1210~n)^{\frac{1}{2}}\eqno(15)$$

\noindent This formula  gives the following values for the exponents:
$\xi_{1/4}\approx 0.1131$;  $\xi_{1/2}\approx 0.2171$;  $\xi_{3/4}\approx 0.3140$;
$\xi_{1}\approx 0.4046$;  $\xi_{2}=0.7257$; $\xi_{4}=1.2433$; $\xi_{5}\approx 1.4644$;
$\xi_{6}\approx 1.6684$.
It is interesting that, strictly speaking, the scalings of all moments,  
including those with
$n<1$, are anomalous and cannot be obtained from dimensional considerations.  It is 
possible  to formally  continue the expression (15) into the interval $n<0$.  We can see that
the moments $S_{n}$  with $n<n_{c}=-\frac{1.3958}{1.1210}\approx -1.2451$ do not exist.
Since the PDF $P(\Delta u, r)\neq 0$ at the origin $\Delta u=0$, it is expected that the 
moments $S_{n}(r)$ with $n<-1$ diverge. Thus, the fact that the relation  (15) gives
the critical moment-order
$n_{c}\approx -1.24$, close to $n_{c}=-1$, indicate  that formula (15) might give 
close-to-
correct values of the scaling exponents $\xi_{n}$ for $n>n_{c}$,  which are 
not too far from  critical $n_{c}=-1$.  We have:
$\xi_{-1/4}\approx -0.126;~\xi_{-1/2}\approx -0.269;~ 
\xi_{-3/4}\approx -0.438;~\xi_{-1}\approx -0.660~$.

Let us now consider the expression for the dissipation term $D$ which is proportional to
$\nu_{0}\rightarrow 0$. The eddy viscosity approximation limits the renormalized
perturbation expansion in powers of the non-linearity by  one-loop contributions 
only. The technical details of the procedure vere described in Refs. [10], [11], [12].
We are interested in the  correlation function, similar to $D$:

$$D_{1}=\nu_{0}~<u(q_{1})....q_{i}^{2}u_(q_{i})...u(k-q_{1}....-q_{n-1})>\eqno(16)$$

\noindent where $i=1;~ 2;.....n-1;$ there is one contribution involving 
$q^{2}_{n}=(k-q_{1}....-q_{n-1})^{2}$.  The expansion is generated by the iteration procedure involving
the Navier-Stokes equations (5) or (6) (see Refs. [10], [11]).  In the zero -loop 
approximation we have 

$$D_{1}^{0}\approx n{\cal E} S_{n-2}(r)$$

\noindent where the $O(1)$ dissipation rate ${\cal E}=\nu_{0}\int k^{2}E(k)dk$. 
This term is obtained from (16) in the limit of the ``free momentum'' 
$k\rightarrow 0$.
Substituting this expression into the equation of motion (9) we find that, for 
the solution 
found above,   balance is impossible 
 since $-1+\frac{\xi_{2}}{2}+\xi_{n}\neq \xi_{n-2}$.  Moreover,  it is easy to see that
the zero-loop contributions to the expression for $D$, we are interested in,  
cancel in the 
limit $k\rightarrow 0$.
In the one-loop approximation we can have $n$  u.v.- divergent
contributions of the order $\nu_{0}S_{n}k^{2}_{d}$ and only one term coming from
the $k$-dependent $q_{n}$ in relation (16) (here $k$ is a free ``momentum''):

$$D_{1}^{2}\propto \nu(r)\frac{dS_{n}}{dr^{2}}\propto \frac{\nu(r)S_{n}}{r^{2}}\eqno(17)$$

\noindent where the effective (eddy) viscosity $\nu(r)=O(r^{1 +\frac{\xi_{2}}{2}})$. 
Substituting (17) into (9) and fixing  the  proportionality coefficient to satisfy 
Kolmogorov's  $4/5$ law (1) we derive the equation (13) used for evaluation of 
the exponents $\xi_{n}$. The u.v.-divergent contributions
 are assumed  to cancel for the eddy viscosity approximation to work.  
This fact was proved to be correct for the  model 
case of a simple effective forcing function in Ref. [12]. It was shown  
that this cancellation is the result of overall energy balance that requiers  
$<{\bf S\cdot u}>={\cal E}$.
In general, the rigorous demonstration that it is so 
is a difficult  task.   All we can say  now is that the right side of equation (13) 
is consistent with the eddy viscosity approximation.  

The fact that eddy viscosity works so well for the description of quite complex 
engineering flows was known for many years in the engineering community. For 
these many years,  the eddy viscosity  concept was treated with suspicion or 
even contempt by many physisists
due to the long-
standing belief that,  being a result of  one- loop closure, it cannot be 
useful  for the 
description of the ``non-perturbative'' intermittency of  turbulence.  
Recently, it has been 
shown 
in direct numerical experiments on the two-time, two-point correlation function
in  one-dimensional 
Kolmogorov turbulence for  a forced Burgers equation that  eddy viscosity works well
to descibe the small scale 
dynamics in this complex and strongly intermittent system [13]. The work of 
Shraiman and Siggia [8], 
as well as other  recent contributions to the theory of a passive scalar [2]-[4] and [7],
also based on the eddy diffusivity,  showed that anomalous scaling can be derived in the 
eddy diffusivity approximation, provided the zero modes are treated with due respect.
(In the problem of a passive scalar advected by a white-in-time random velocity
the eddy diffusivity is an exact consequence of the equations of motion).

 The  scaling exponents, calculated by  the present anlysis, are
compared with the results of physical experiments (Refs.  [14]-[16]) in  Table I. 
The measurement of the two-point correlation functions is very difficult. That is 
why usually the single-point, two-time correlations functions $F(\tau)=<u(x,t)u(x,t+\tau)>$
are measured and a simple space-time tranformation $r=U\tau$ is used for 
the interpretation of the data in terms of  spatial correlation functions. 
This transformation is called Taylor frozen turbulence hypothesis,  which is accurate 
when the ratio of the fluctuating $u_{rms}$
and mean $U$ velocities is very large $a=U/u_{rms}>>1$. This criterion is well 
satisfied 
in the wall-bounded  flows (boundary layers, pipes, ducts etc ),  while in 
the jets, mixing 
laers and other open flows $a=O(1)$.  The experimental data used in the Table were  
measured in the atmospheric boundary layer [14] with $a\approx 30$,  in the flow 
between counterrotating 
disks [15] with $U=0$ and in the turbulent jet [16].  To account for the space-time 
relation  in an accurate way, 
the authors of Ref. [15] introduced a Lagrangian-like transformation,  based on the 
idea of the ``local Taylor hypothesis'',  which can be accurate even when $a=0$. 
The data of Ref. [16] were collected in the jet flow using a novel nonintrusive 
optical technique which enables one to measure directly multi-point spatial 
correlation functions. 
Due to the experimental uncertainty, we have 
avoided comparison with  the data on the two-time correlation functions,  
based on the traditional Taylor hypothesis, obtained in the open flows.

The fact that the scaling exponents evaluated in this work agree so well with  
experimental data
is   additional evidence that the eddy viscosity approximation is  much more 
powerfull than may have  expected  for  a simple  one-loop theory.  
\noindent The eddy viscosity approximation,  applied to the inertial range dynamics,
 accounts for  pressure fluctuations only 
in the numerical value of the factor $\Gamma$ in the eddy viscosity definition [10], [12].
This is not good enough  for the correct representation of the effects  ifluenced by  
large-scale
velocity fluctuations,  which dominate very high-order  moments of velocity 
differences (vortex filaments etc).
Moreover, the equation (13) takes into account only the first dissipative  anomaly
given by (1). It is clear the anomalies, describing constant or close-to-constant 
fluxes
of other flow characteristics,  like $K=u_{i}u_{i}$,   can influence the properties of
high- order moments.  Since the theory does not account  for these 
effects, 
it cannot be used for evaluation of the exponents when $n$ is large enough.  To conclude, 
we would like to mention recent multifractal model by She and Leveque [17],  leading 
to a different  expression for 
the scaling exponents $\xi_{n}$ which are in a good agreement with experimental data 
for $n>0$.  The relation between the equation (13) and formula (15), derived here, and 
the She-Leveque theory is not understood. 

\noindent I would like to thank K.R.Sreenivasan for communicating his 
exprerimental findings on the 
scaling exponents of low-order structure functions prior to publication. Helpful
discussions with R.H. Kraichnan, M. Nelkin and S. Orszag are aknowledged. This work was
supported in parts by ONR, AFOSR and ARPA.

\begin{tabular}{|r@{}|r@{}|r@{}|r@{}|r@{}|}
\hline
\multicolumn{1}{|c|}{n} &
\multicolumn{1}{c|}{$\xi_n$,calc}&
\multicolumn{1}{c|}{$\xi_n$,[14]}&
\multicolumn{1}{c|}{$\xi_n$,[15]}&
\multicolumn{1}{c|}{$\xi_n$,[16]}
\\ \hline \hline
0.1 & 0.0466 &$ 0.043 \pm 0.006$ &           &                 \\ \hline
0.2 & 0.0913 &$ 0.083 \pm 0.010$ &           &                 \\ \hline
0.3 & 0.1346 &$ 0.123 \pm 0.110$ &           &                 \\ \hline
0.5 & 0.2171 &$ 0.200 \pm 0.150$ &           &                 \\ \hline
1   & 0.4046 &$ 0.384 \pm 0.023$ & 0.40      &                 \\ \hline
1.5 & 0.5727 &$ 0.555 \pm 0.024$ &           &                 \\ \hline
2   & 0.7257 &$ 0.714 \pm 0.025$ & 0.71      & $0.70 \pm 0.01$ \\ \hline
4   & 1.2433 &  1.21             & 1.24      & $1.28 \pm 0.03$ \\ \hline
5   & 1.4644 &  1.53             & 1.48      & $1.50 \pm 0.05$ \\ \hline
6   & 1.6684 &  1.66             & 1.69      & $1.75 \pm 0.10$ \\ \hline
8   & 2.0378 &  2.05             &           &                 \\ \hline
10  & 2.3690 &  2.38             &           &                 \\ \hline

\end{tabular}
\\
\\

{\bf references}
\\
1. A.N.Kolmogorov, Dokl. Akad. Nauk SSSR, {\bf 30}, 299 (1941)
\\
2. R.H.Kraichnan, Phys.Rev.Lett., {\bf 72}, 1016 ( 1994)
\\
3. K. Gawedzki and A. Kupiainen, Phy.Rev.Lett., {\bf 75}, 3608  (1995)
\\
4. M. Chertkov, G. Falkovich, I. Kolokolov  and V. Lebedev, 
Phys.Rev.E, {\bf 51}, 4924 (1995)
\\
5. Ya. G. Sinai and V.Yakhot, 1988, unpublished
\\
6. R. H. Kraichnan, Phys. Fluids, {\bf 11}, 945 (1968)
\\
7. V. Yakhot, Phys.Rev.E,  1996 (in press)
\\
8. B. Shraiman and E. Siggia, C.R.Sci. (Paris) {\bf 321}, 275 (1995)
\\
9. R.H. Kraichnan, Phys. Fluids, {\bf 9}, 1728 (1966)
\\
10. V. Yakhot and S. A. Orszag, Phys. Rev. Lett., {\bf 57}, 1722 (1986)
\\
11. V. Yakhot Orszag, J. Sci. Comp., {\bf 1}, 3 (1986)
\\
12.  V. Yakhot  and L.M. Smith, J.Sci.Comp., {\bf 7}, 35 (1992)
\\
13. A. Chekhlov, V. Yakhot, Phys. Rev. E, {\bf 51}, R2739 (1995)
\\
14. The low-oredr moments $S_{n\leq 2}$ were measured recently by 
B. Dhruva and K.R. Sreenivasan, private communication,
The moments with $n>2$, quoted in the Table,  are from
G. Stolovitzky, K.R. Sreenivasan and Juneja, Phys. Rev. E, {\bf 48}, R3217 (1993)
\\
15. J.-F. Pinton and  R. Labbe, J.Phys. II France, {\bf 4}, 1461 (1994)
\\
16. A.Nullez, U. Frisch, R. Miles, W. Lempert, private communication
\\
17. Z.-S. She and E. Leveque, Phys.Rev.Lett., {\bf 72}, 336 (1994)
\end{document}